\documentclass[]{emulateapj}

\usepackage{epsfig,amsmath,amssymb,amsfonts,mathrsfs,latexsym,graphicx}
\usepackage{hyperref}

\begin{document}

\title{Low-frequency QPO from the 11~Hz accreting pulsar in Terzan 5:\\
  not frame dragging }

\author{D. Altamirano\altaffilmark{1}, A. Ingram\altaffilmark{2},
  M. van der Klis\altaffilmark{1}, R. Wijnands\altaffilmark{1},
  M. Linares\altaffilmark{3} \& J. Homan\altaffilmark{3}}

\altaffiltext{1}{Email: d.altamirano@uva.nl ; Astronomical Institute,
  ``Anton Pannekoek'', University of Amsterdam, Science Park 904,
  1098XH, Amsterdam, The Netherlands.}

\altaffiltext{2}{Department of Physics, University of Durham, South
  Road, Durham DH1 3LE, UK.}

\altaffiltext{3}{Massachusetts Institute of Technology - Kavli Institute for
  Astrophysics and Space Research, Cambridge, MA 02139, USA.}

\begin{abstract}

We report on 6 RXTE observations taken during the 2010 outburst of the
11 Hz accreting pulsar IGR~J17480--2446 located in the globular
cluster Terzan 5.
During these observations we find power spectra which resemble those
seen in Z-type high-luminosity neutron star low-mass X-ray binaries,
with a quasi-periodic oscillation (QPO) in the 35--50 Hz range
simultaneous with a kHz QPO and broad band noise.
Using well known frequency-frequency correlations, we identify the
35--50 Hz QPOs as the horizontal branch oscillations (HBO), which were
previously suggested to be due to Lense-Thirring precession.
As IGR~J17480--2446 spins more than an order of magnitude more slowly
than any of the other neutron stars where these QPOs were found, this
QPO can not be explained by frame dragging.
By extension, this casts doubt on the Lense--Thirring precession model
for other low-frequency QPOs in neutron-star and perhaps even
black-hole systems.

\end{abstract}

\section{Introduction}\label{sec:intro}

One of the strongest motivations for studying low-mass X-ray binaries
(LMXBs) has been the aim to use these systems as probes of fundamental
physics.
A possible tool for this is provided by the quasi-periodic
oscillations (QPOs) in the X-ray light curves of LMXBs.
These QPOs have now been observed in many LMXBs containing either
neutron stars (NSs) or black holes (BHs), and are usually detected
with characteristic frequencies between $\sim$1 mHz and $\sim$1 kHz.
The QPO frequency, coherence and amplitude usually correlate with the
source spectral states and/or X-ray luminosity, observables probably
set by the geometry and dynamics of the accretion flow \citep[see,
  e.g.,][for a review]{Vanderklis06}, supporting the idea that QPOs
can be used as probes of the flow of matter in strong-field gravity
and hence of fundamental physics.

Persistent NS systems have been generally divided into low-luminosity
(``4U'') and high-luminosity (``GX'') atoll sources, and the always
high-luminosity Z-sources, based not only on luminosity, but also on
the tracks they trace out in color-color (CCD) and hardness-intensity
diagrams (HID) and on their correlated rapid X-ray variability
\citep{Hasinger89}.
There is no similar subdivision for BH systems.

A number of QPOs and broad-band variability components are often
present simultaneously in the power spectra of the X-ray light curves
of these systems.
These power spectra can be fully described with a phenomenological
model comprising several Lorentzian components, where each component
is generally labeled by its characteristic frequency
\citep[e.g.][]{Belloni02}.
Recently, more physically-motivated models based on Lorentzian mass
accretion rate fluctuation spectra in the disk have begun to be
explored \citep[see][]{Ingram09, Ingram10, Ingram11, Ingram12}.
When the multi-Lorentzian model is used, each component is called
$L_i$, and its characteristic frequency is $\nu_i$, where $i$ is an
identifying symbol \citep[e.g.][]{Altamirano08}.

\begin{figure} 
\centering
\resizebox{1\columnwidth}{!}{\rotatebox{0}{\includegraphics[clip]{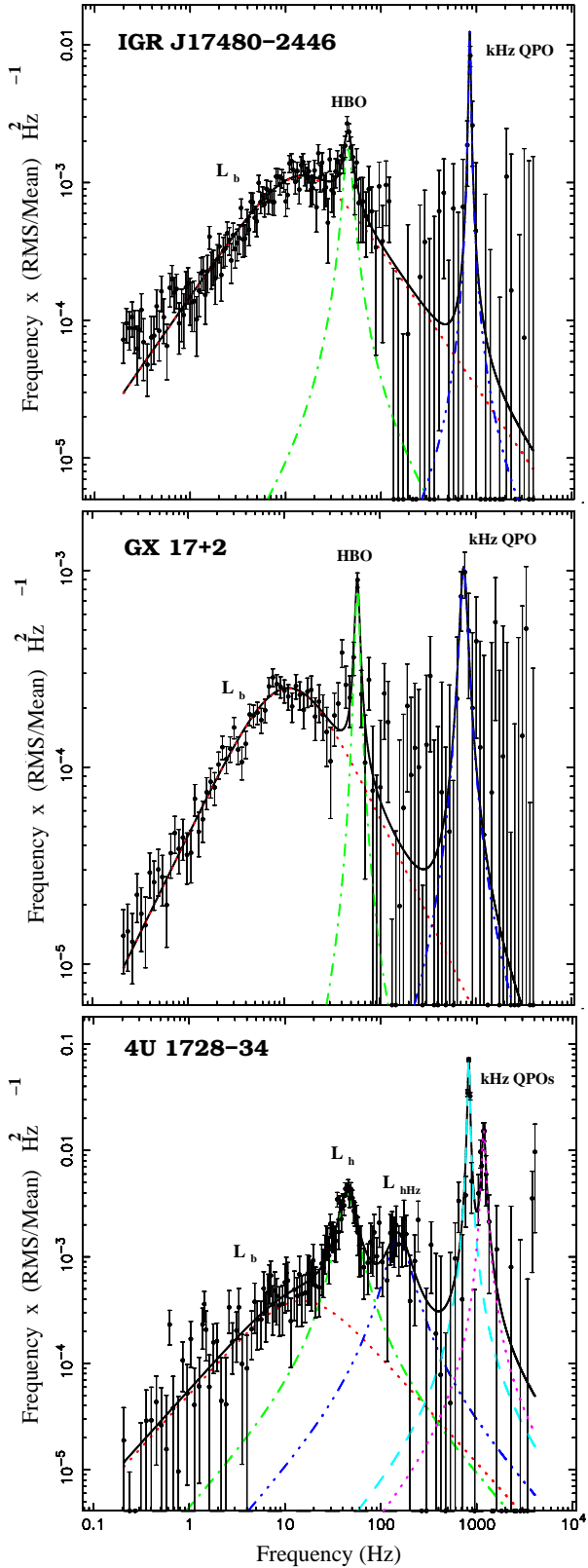}}}
\caption{\textit{Top:} Representative power spectrum of
  IGR~J17480--2446 (ObsID:95437-01-09-00) during which we detect a
  low-frequency and a single kHz QPO. \textit{Middle} and
  \textit{Bottom} panels show power spectra for the Z-source GX17+2
  (ObsID:20053-03-02-010) and the atoll source 4U1728--34 (ObsID:20083-01-04-00), respectively.
All were made using 2--60 keV RXTE data.
Power-spectral components are labeled in each panel.}
\label{fig:pds}
\end{figure}

Power spectral features in atoll sources include broad components
known as the break component ($L_b$), the hump $L_{h}$, the $L_{\ell
  ow}$ and narrower components, or QPOs, such as the low-frequency QPO
($L_{LF}$) and the upper and lower kHz QPOs ($L_u$ and $L_{\ell}$,
respectively).
For Z-sources, low-frequency QPOs have been labeled differently: the
horizontal-branch, normal-branch and flaring-branch oscillations
(HBOs, NBOs and FBOs, respectively) depending on which spectral state
(or ``branch'' in the HID) of the source they are most prominent. In
Figure~\ref{fig:pds} we show examples; for a more detailed
description, we refer the reader to the review by
\citet{Vanderklis06}.
In addition to the break component (and some other broad-band
components like $L_{\ell ow}$), BHs generally show three main types of
low-frequency QPOs \citep[Types A, B and C, e.g.,][]{Casella05} and,
in a few cases, high-frequency QPOs \citep[with frequencies between
  70 Hz and 450 Hz, e.g.,][]{Remillard06,Belloni12}.

Similarities in the morphology of their power spectra suggest that
same variability components are present in both NS and BHs
\citep[e.g., ][]{Miyamoto93,Vanderklis94,Olive98,Belloni02,Linares07}.
Correlations between the frequencies of some of these power-spectral
components, and similarities in their characteristics confirm this
suggestion, providing intriguing links between NSs and BHs.
In particular, it has often been suggested that the HBOs in Z-sources
are similar to the LF QPO and hump component in atoll sources
\citep[e.g.][]{Straaten03} and in BHs \citep[e.g.][]{Casella05}.

Studies have been done on the relation between the frequencies of
various variability components in compact objects. The two best known
frequency-frequency correlations which involve both NSs and BHs are
those known as WK \citep{Wijnands99a} and PBK \citep{Psaltis99b}
relations.
The WK relation links the frequencies of the well identified $L_b$
noise component in atoll sources and BHs, to the $L_h$/$L_{LF}$ LF-QPO
in atoll and BH sources\footnote{As noted by \citet{Belloni02},
  \citet{Klein08} and \citet{Straaten03}, $L_{LF}$ and $L_h$ are often
  close in frequency, separated by no more than a few
  Hz.}. \citet{Wijnands99a} showed that the frequencies are well
correlated over 3 orders of magnitude. In addition, these authors
found that the HBOs in Z-sources also show the same relation with
$L_b$, albeit slightly above the atoll/BH track.
The PBK relation links the second highest frequency observed in NS
($L_\ell$ in the high-luminosity soft state and $L_{\ell ow}$ in the
low-luminosity hard state) and BHs ($L_{\ell ow}$ in the low-hard
state) on one hand, and low-frequency QPOs on the other
\citep[e.g.,][]{Belloni02}.
This relation spans nearly three decades in frequency \citep[and even
  a larger range if one considers QPOs in white dwarfs systems, see,
  e.g.,][]{Warner02}. However, as \citet{Psaltis99b} note,
although the correlation is very suggestive, the relation between
components is not conclusive as it combines features with different
coherence and amplitudes from different sources.

The WK and PBK relations suggest that physically similar (or
identical) phenomena set the frequencies observed in BHs and low- and
high-luminosity NSs. If true, then the mechanism that sets their
frequency must arise in the accretion disk, i.e., it cannot depend on
a solid surface \citep[e.g., ][]{Wagoner01,Rezzolla03}.

\subsection{Relativistic orbital motion and the identification of Lense-Thirring precession}\label{sec:LT}

Orbits tilted relative to the equatorial plane of a central spinning
object show relativistic nodal precession due to frame dragging.  In
the Lense-Thirring (LT) approximation \citep[weak field and low spin,
  see][]{Lense18} the precession frequency is given by $\nu_{\rm
  LT}=GJ/\pi c^2r^3$ where $J$ is the angular momentum of the central
object and $r$ the orbital radius.

In this approximation the LT precession frequency around a spinning
NS can be written as

\begin{eqnarray}
\begin{split}
\nu_{{\rm LT}} = 13.2{\rm
  Hz} \ I_{45} \left(\frac{M}{M_\odot}\right)^{-1} \left(\frac{\nu_\phi}{1000{\rm
    Hz}} \right)^2 \left(\frac{\nu_{{\rm spin}}}{300{\rm Hz}} \right)
\end{split}
\end{eqnarray}

where $\nu_\phi$ is the orbital frequency, $I_{45}$ the moment of
inertia of the NS in units of 10$^{45}$ g cm$^2$, $M$ its mass, and
$\nu_{{\rm spin}}$ its spin frequency \citep{Stella98}.

\citet{Stella98} proposed that the low-frequency QPOs (HBOs and given
WK and PBK, also the $L_{LF}$ and/or $L_h$) in NS systems represent LT
precession of the orbit whose general-relativistic orbital frequency
$\nu_\phi$ is given by the upper kHz QPO frequency
$\nu_u$. Furthermore, \citet{Stella99a} proposed that via the
correlations identified by PBK this interpretation can be extended to
LF QPOs seen in BH systems.
For the NS case, the predicted and observed frequencies do not match
directly: as \citet{Stella98} note, for reasonable values of
$I_{45}/(M/M_\odot)$ and $\nu_{{\rm spin}}$, the predicted
precession frequencies were a factor of $\sim$2 lower than the
observed ones \citep[or even a larger factor, e.g.][]{Jonker98,
  Morsink99, Psaltis99c, Jonker00c}.
This discrepancy was then interpreted by assuming that the modulation
can be produced at twice the LT precession frequency \cite[see
  discussion in][]{Stella98}, which is not unreasonable given the
bilateral symmetry inherent in the geometry of a tilted precessing
orbit.

Based on the works of \citet{Liu02} and \citet{Fragile07},
\citet{Ingram10} recently were able to naturally explain the factor of
$\sim$2 discrepancy by relaxing the test-particle assumption, and
considering LT precession of a geometrically thick inner accretion
flow, where a thin truncated outer disk remains stationary but the
inner flow precesses as a solid body.

One of the ingredients in all the above works is that the spin
frequency of the NS is in the range 200--600 Hz.  Most of the
$\nu_{{\rm spin}}$ measurements in NS-LMXBs are in that range
\citep{Patruno12a}.
The recent discovery of a new X-ray transient in the globular cluster
Terzan 5 containing an 11 Hz pulsar allows us to examine the frame
dragging model in a new regime.

\subsection{Testing LT precession models with IGR~J17480--2446}

IGR~J17480--2446  was discovered in the globular cluster Terzan 5 on
October 10th, 2010, with INTEGRAL.
Initial RXTE observations revealed an 11 Hz pulsar
\citep{Strohmayer10a} in a 21.3 hr orbital period binary
\citep{Papitto11}.
This spin is slow compared to the 185--650~Hz known spin of the
low-magnetic ($\sim$10$^8$ Gauss) field NS-LMXBs, yet (much) faster
than that of the ($5\times10^{-5}-2$)~Hz pulsars, which exhibit
various characteristics indicating a strong ($\gtrsim10^{11-12}$ Gauss)
magnetic field strength \citep[e.g.,][for a review]{Patruno12a}.
The magnetic field of IGR~J17480--2446 may be intermediate in strength
($\sim$$10^{10}$ Gauss, e.g., \citealt{Cavecchi11}, \citealt{Papitto11}
and \citealt{Patruno12}).

Near the outburst peak, at about half the Eddington luminosity
\citep[e.g.][]{Chakraborty11a,Linares12}, IGR~J17480--2446 showed
X-ray spectral and variability behavior typical of Z sources
\citep{Altamirano10h}, with simultaneous broad-band noise, a QPO at
$\sim$48 Hz, and a kHz QPO at $\sim$815 Hz.
These power-spectral components resemble those seen in other NS
systems, where the $\sim$48 Hz QPOs are the ones identified with LT
precession.

\section{Observations and data analysis}\label{sec:observations}

We use data from the RXTE Proportional Counter Array \citep[PCA; for
  instrument information see][]{Jahoda06}. There were 48
pointed observations, each consisting of a fraction of one to several
entire satellite orbits.
For the timing analysis we used the Event mode
E\_125us\_64M\_0\_1s.  Leahy-normalized power density spectra
\citep{Leahy83} were constructed using data segments of 128 seconds
and 1/8192~s time bins.
All frequency bins between 11.02 and 11.07 Hz were removed to get rid
of the pulsar spike.
No background or deadtime corrections were performed prior to the
calculation of the power spectra.  We averaged the power spectra per
orbit and per observation, subtracted a modeled Poisson noise spectrum
\citep{Zhang95} and converted the resulting power spectra to squared
fractional rms \citep{Vanderklis95b}.

For a detailed analysis of the power and energy spectral evolution
along the outburst, we refer the reader to \citet{Altamirano12prep}
and \citet{Barret12}. In the current work, we concentrate on the 6
observations where the low-frequency QPOs and the kHz QPOs were
detected.

\section{Results}\label{sec:results}

We found LF QPOs with frequencies between $\sim$35 Hz and $\sim$50 Hz
in 6 observations (9 independent satellite orbits). This QPO was not
always present during the whole observation.
kHz QPOs with frequencies between $\sim$800 Hz and $920$ Hz
\citep{Altamirano12prep, Barret12} were detected in 5 observations,
simultaneously with the low-frequency QPOs \citep{Altamirano12prep}.
In the top panel of Figure~\ref{fig:pds} we show a representative
power spectrum of IGR~J17480--2446 where we detect a broad feature
(zero-centered Lorentzian) with a characteristic frequency of
$14.8\pm0.8$ Hz, a QPO at $45.0\pm0.7$ Hz and a kHz QPO at $851\pm4$
Hz.
The fractional amplitude of the 45 Hz QPO increases with energy, from
undetected in the 2--5 keV (with a 3$\sigma$ upper limit of 2.9\% rms)
to $7.3\pm1.3$\% rms at $\sim$20 keV.
The statistics are not sufficient to study the QPO phase-lags.
The overall power spectral shape resembles that observed in some Z
sources, and to a lesser extent, those observed in atoll sources
\citep[middle and bottom panels in Figure~\ref{fig:pds}, respectively;
  see also][]{Altamirano12prep}.
The frequency range, the quality factor and rms amplitude of the
35-50~Hz QPOs, together with the overall power-spectral shape and the
position where the QPOs occur in the HID \citep{Altamirano12prep}, are
all consistent with these QPOs being the HBO in Z sources.
The kHz QPO can be identified as either the upper or the lower kHz QPO
\citep{Altamirano12prep,Barret12}.

\begin{figure} 
\centering
\resizebox{1\columnwidth}{!}{\rotatebox{0}{\includegraphics[clip]{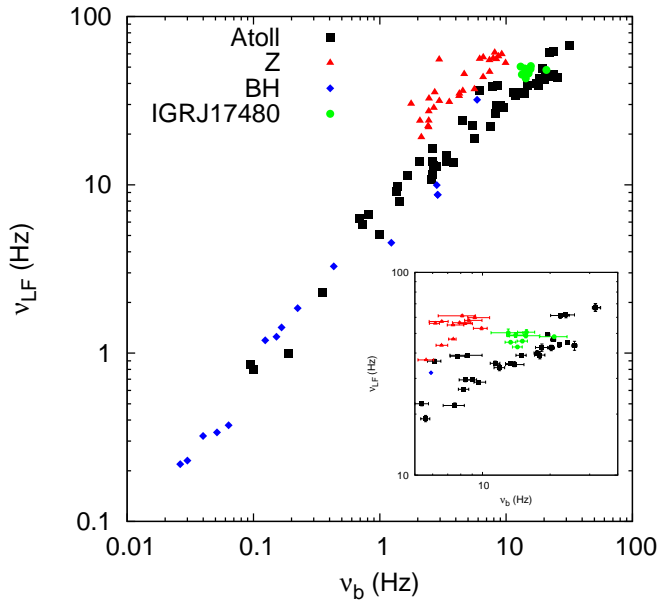}}}
\caption{WK relation after \citet{Wijnands99a}. Black squares, red
  triangles and blue diamonds are data from atoll sources, Z sources,
  and BHs, respectively. Green circles are from
  IGR~J17480--2446. Inset shows a zoom-in.  No errors are shown in the
  main figure for clarity.}
\label{fig:WK}
\end{figure}

We further tested our identifications using the WK relation, as the
break component $L_b$ and the low-frequency QPO are easy to identify.
In Figure~\ref{fig:WK} we plot the data from \citet{Wijnands99a}. Our
points for IGR~J17480--2446 are on the main relation, hugging the
upper envelope of the atoll sources (black squares), but still below
the Z-sources (red triangles). 
(Slight differences in how $\nu_b$ is measured will affect the values
by $<15$\% \citep{Belloni02} and hence are immaterial to our
conclusions.)

\begin{figure} 
\centering
\resizebox{1\columnwidth}{!}{\rotatebox{0}{\includegraphics[clip]{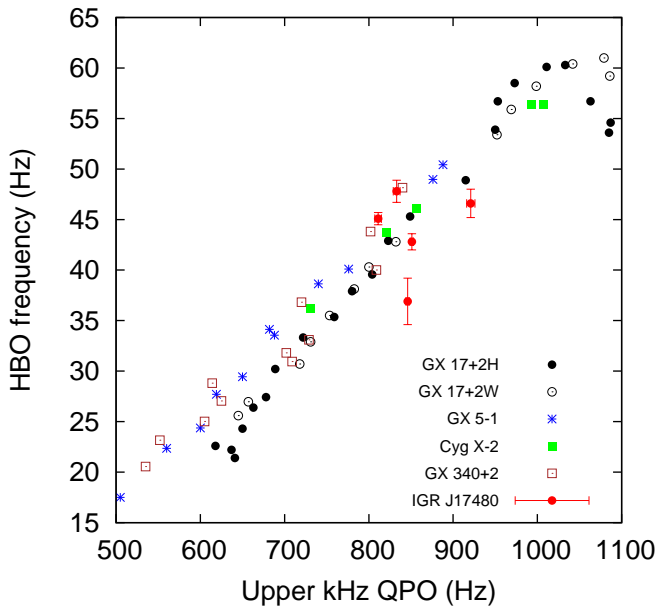}}}
\caption{HBO vs. upper kHz QPO frequency for different Z sources
  \citep[data from][]{Altamirano12prep}. Red circles represent our
  simultaneous detections of the HBO and kHz QPO assuming that the
  single kHz QPO is the upper one, or that $\Delta\nu$ is of the order
  of the spin frequency of IGR~J17480--2446 (i.e. 11 Hz). For clarity,
  we show only error bars for IGR~J17480--2446 data. }
\label{fig:gx17}
\end{figure}

As discussed by \citet{Mendez07}, the available results on NS kHz QPOs
are inconclusive on whether $\Delta\nu=\nu_u-\nu_\ell$ is related
to the spin frequency of the NS as $\Delta\nu=\nu_{{\rm spin}}$, or
as $\Delta\nu=\nu_{{\rm spin}}/2$, or it is independent of
$\nu_{{\rm spin}}$ and close on average to $\sim$300 Hz.
In Figure~\ref{fig:gx17} we compare IGR~J17480--2446's QPO frequencies
with those of the HBO and upper kHz QPO in other Z sources \citep[see
][]{Altamirano12prep}. The data are suggestive of the correct
identification of the different components assuming that either the
kHz QPO we detect is the upper, or it is the lower and
$\Delta\nu\ll$300 Hz.
The PBK relation (not shown) uses the frequency of the lower kHz
QPO. In this case, the data of IGR~J17480--2446 are also on the main
correlation if we assume that we are detecting the upper kHz QPO, and
that $\Delta\nu$ is around 300 Hz.

\section{Discussion}

In this paper we report on 6 observations of the 11 Hz accreting
pulsar IGR~J17480--2446 in which we find power spectra which resemble
those seen in Z-type high-luminosity NS-LMXBs.
The low-frequency QPOs are in the 35-50 Hz range, while the kHz QPOs
are in the 800--920 Hz range.
Based on the power-spectral characteristics, where the QPOs occur in
the HID \citep{Altamirano12prep}, and comparison with
frequency-frequency correlations, we identify the low-frequency QPO as
the HBO. Our results suggest that the kHz QPO is the upper one,
although the results are not conclusive.
For a more detailed discussion about the identification of the kHz
QPO, we refer the reader to \citet{Altamirano12prep}. In this letter,
we discuss our results mainly in the context of Lense-Thirring
precession.
Comparison of our results with other proposed models for low-frequency
QPOs will be reported elsewhere.

\subsection{Lense-Thirring precession}

The LT precession frequency for a test particle as
introduced in Section~\ref{sec:LT} depends on the NS spin, its
moment of inertia and mass, and on the orbital frequency.
For IGR~J17480--2446 $\nu_{{\rm spin}}=11$ Hz. Depending on whether
the NS equation of state (EoS) is soft or stiff, respectively,
$I_{45}/(M/M_\odot)$ could be between 0.5 and 2
\citep{Stella98}. Assuming $I_{45}/(M/M_\odot)<2$, we obtain $\nu_{\rm
  LT}\lesssim0.97 \ {\rm Hz}\times(\nu_\phi/1000 \ {\rm Hz})^2$.
If we associate the 800--920 Hz kHz QPO frequency with $\nu_\phi$,
then $\nu_{\rm LT}<0.82$ Hz, i.e. much lower than the 35-50 Hz QPO
we observe.

\begin{figure} 
\centering
\resizebox{1\columnwidth}{!}{\rotatebox{0}{\includegraphics[clip]{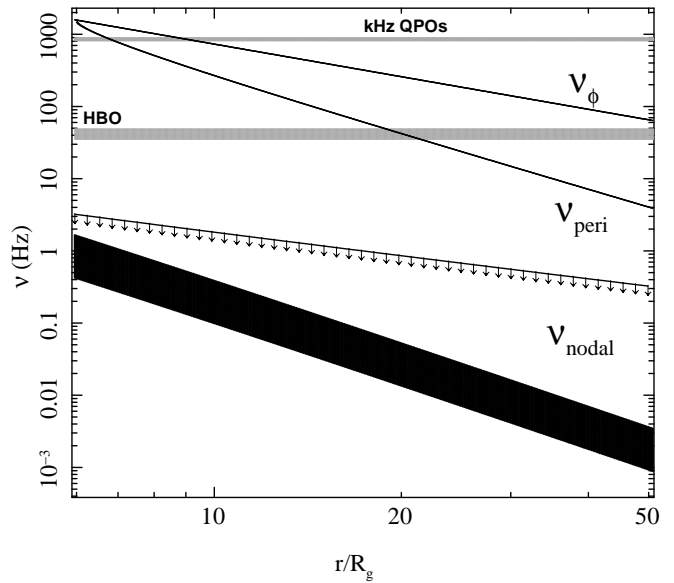}}}
\caption{General-relativistic orbital frequency $\nu_\phi$, periastron
  precession frequency $\nu_{\rm peri}$ and nodal precession frequency
  $\nu_{{\rm nodal}}$ as a function of orbital radius $r$ for an 11 Hz
  pulsar calculated from the expressions in \citet{Merloni99}.
  Horizontal shaded bands at 35-50~Hz and 800-900~Hz indicate the
  observed HBO and kHz QPO frequencies.
Black shaded area at bottom marks $\nu_{{\rm nodal}}$ for prograde
orbits and $R=6R_g$; upper and lower bounds are practically
independent of mass and correspond to $\beta=I/(MR^2)$ equal 2/3 and
1/6.  For $M=1.4M_\odot$ these bounds correspond to $I_{\rm
  45}/(M/M_\odot)$ $\sim$2 and $\sim$0.5, respectively.
The curve with arrows is an upper limit to $\nu_{{\rm nodal}}$ for
extreme assumptions maximizing this frequency: retrograde orbital
motion, $\beta=2/3$, and NS radius equal to the orbital radius
($R=r$).  Even with extreme assumptions theoretical nodal precession
frequencies remain far below the observed HBO frequencies.
At this low spin $\nu_\phi$ and $\nu_{\rm peri}$ (both for prograde
orbits) are practically independent of $R$ and $\beta$ but they do
depend on mass; curves are for 1.4$M_\odot$.  These curves can match
the observed frequencies, but not at the same orbital radius.}
\label{fig:freq}
\end{figure}

We can take a step further and estimate the relativistic nodal
precession frequency independently of the nature of the kHz QPOs. The
LT precession frequency (Section~\ref{sec:LT}) can be written as
$\nu_{\rm LT}=2G\nu_{\rm spin}I/c^2r^3$.  Writing the NS
moment of inertia as $I=\beta~MR^2$ where
$\beta$ is a dimensionless constant, we have $\nu_{\rm LT}=2\beta
R_gR^2\nu_{\rm spin}/r^3$, where the gravitational radius
$R_g\equiv GM/c^2$.
The highest possible precession frequency occurs at the NS surface
($r=R$), giving $\nu_{LT}<2\beta\nu_{\rm spin}(R_g/R)$. The maximum
value for $I$ is reached when a hollow sphere is assumed
($\beta=2/3$). This is, of course, not a realistic assumption for a
NS, we employ it merely as a hard upper limit for $\beta$. The true
nature of the space-time external to such a shell is therefore
irrelevant to our estimate. Assuming the NS radius to coincide with
its own innermost stable circular orbit, which for such low spin is
$r_{\rm ISCO}=6R_g$ to a very good approximation, we find $\nu_{\rm
  LT}<2.44$Hz.
Higher order terms in a Kerr metric \citep[e.g.,][]{Merloni99},
deviations of the metric from Kerr related to the NS structure
\citep[e.g.,][]{Morsink99}, and arbitrarily inclined orbits
\citep[e.g.,]{Sibgatullin02} could affect this upper limit by a factor
of a few at most, and some of these effects actually lower it.

Considering that the entire inner disk precesses as a solid body between
an inner and outer radius \citep[e.g.,][]{Ingram10} does not solve the
discrepancy either, as in their description the entire disk cannot
precess faster than a test particle at the inner radius.
Our conclusion is that the 35--50 Hz QPO in IGR~J17480--2446
can not be explained by frame dragging.

Figure~\ref{fig:freq} shows $\nu_\phi$, $\nu_{{\rm peri}}$ and
$\nu_{{\rm nodal}}$ as a function of radius $r$ for a point mass in
the Kerr metric in an infinitesimally tilted and eccentric orbit
\citep{Merloni99}.
In particular it shows a range of solutions for $\nu_{\rm nodal}$, with
the upper limit resulting from assuming retrograde orbits and $R=r$.
We note that theoretically the periastron precession frequency
$\nu_{\rm peri}$ could in this system be identified with the QPOs at
35--50 Hz, but at a much larger radius ($r\sim$20$R_g$) than that
where $\nu_\phi$ is produced ($r\sim$9$R_g$). However, this would be
entirely ad-hoc and would not work for the fast-spin NSs, where
$\nu_{\rm peri}$ has instead been proposed to be identified with the lower
kHz QPO.

Corrections due to classic precession are always of the order of, or
lower than, $\nu_{\rm nodal}$ (under the assumption that deviations
from a spherical NS are only due to the NS rotation, see
\citealt{Morsink99}; see also \citealt{Laarakkers99}).
\citet{Shirakawa02} found that the NS magnetic field can induce warps
in the inner accretion disk, resulting in a precessing inner flow.
The resulting net precession frequency is set by a combination of
(prograde) LT precession and (retrograde) classical and magnetic
precession. The magnetic precession frequency $\nu_{\rm mag}$ is
independent of $\nu_{\rm spin}$ and scales as $\mu^2$, where $\mu$ is
the magnetic moment. The likely higher $\mu$ in IGR~J17480-2446 as
compared with other NSs \citep{Cavecchi11,Papitto11} suggests the
possibility that magnetic precession could dominate in this system,
opening a different possibility of explaining the LF-QPOs in
IGR~J17480--2446. Why in this scenario IGR~J17480--2446 still conforms
to the WK relation remains to be investigated, as $\nu_{\rm mag}$
depends on $\mu$, the accretion rate, the viscosity of the disk, and
the angle between the magnetic dipole and the spin axis.  If magnetic
precession does dominate in IGR~J17480--2446, then magnetic effects
would be expected to affect observed nodal precession frequencies in
other, presumably lower magnetic-moment LMXBs as well.

In summary: the 35--50 Hz QPOs reported herein are incompatible with
Lense-Thirring (or more generally GR nodal) precession, excluding
frame dragging as the cause of the QPO in IGR J17480-2446. Given the
similarities between the QPOs in IGR~J17480-2446 and other Z-source
systems, we conclude that frame dragging is in doubt as the mechanism
that produces the horizontal-branch oscillations and, possibly, the
hump component and LF-QPOs seen in Atoll sources and BH systems.
While a scenario is conceivable where nodal precession causes all
these QPOs, with magnetic precession dominating in IGR~J17480--2446
and GR precession in the other systems, this requires a coincidence
where systems with different magnetic fields all conform to a
frequency-frequency relation (Figure~\ref{fig:WK}) that previously was
explained assuming no magnetic precession.

\textbf{Acknowledgments:} We are grateful to S. Morsink and P. Uttley
for very insightful discussions. DA and MK acknowledge support from
the International Space Science Institute (ISSI), Team Number 116.
RW is partly supported by an ERC starting grant.
J.H. acknowledges support from a NWO visitors grant.

\end{document}